\documentclass{emulateapj}
\usepackage{subfigure}
 \usepackage[stable]{footmisc}

\begin{document}

\title{Embedded Clusters in the Large Magellanic Cloud using the VISTA Magellanic Clouds Survey \footnotemark[1]} \footnotetext[1]{Based on observations made with VISTA at the Paranal Observatory under program ID 179.B-2003.}
\author{Krista Romita, Elizabeth Lada}
\affil{Department of Astronomy, University of Florida, 211 Bryant Space Science Center Gainesville, FL 32611;\\ k.a.romita@ufl.edu, elada@ufl.edu}
\and
\author{Maria-Rosa Cioni}
\affil{Universit\"{a}t Potsdam, Institut f\'{u}r Physik und Astronomie, Karl-Liebknecht-Str. 24/25, 14476 Potsdam, Germany}
\affil{Leibnitz-Institut f\"{u}r Astrophysik Potsdam, An der Sternwarte 16, 14482 Potsdam, Germany}
\affil{University of Hertfordshire, Physics Astronomy and Mathematics, College Lane, Hatfield AL10 9AB, United Kingdom; \\ mcioni@aip.de}

\begin{abstract}
We present initial results of the first large scale survey of embedded star clusters in molecular clouds in the Large Magellanic Cloud (LMC) using near-infrared (NIR) imaging from the VISTA Magellanic Clouds Survey \citep{vmc}.  We have explored a $\sim$1.65 deg$^2$ area of the LMC, which contains the well-known star-forming region 30 Doradus as well as $\sim$14\% of the galaxy's CO clouds \citep{magma}, and have identified 67 embedded cluster candidates, 45 of which are newly discovered as clusters.   We have determined sizes, luminosities and masses for these embedded clusters, examined the star formation rates (SFRs) of their corresponding molecular clouds, and made a comparison between the LMC and the Milky Way.  Our preliminary results indicate that embedded clusters in the LMC are generally larger, more luminous and more massive than those in the local Milky Way.  We also find that the surface densities of both embedded clusters and molecular clouds is $\sim$3 times higher than in our local environment, the embedded cluster mass surface density is $\sim$40 times higher, the SFR is $\sim$20 times higher, and the star formation efficiency is $\sim$10 times higher.  Despite these differences, the SFRs of the LMC molecular clouds are consistent with the SFR scaling law presented in Lada et al. (2012).  This consistency indicates that while the conditions of embedded cluster formation may vary between environments, the overall process within molecular clouds may be universal.
\end{abstract}

\section{Introduction}
\label{intro}
In the Milky Way, it has been established that the majority of stars form in clusters embedded in clouds of molecular gas \citep{lada92,ladalada,dewit}.  Studying the characteristics and distribution of embedded star clusters is important for our understanding of star formation.  Embedded clusters represent the most recent star forming activity and their properties trace the conditions of their natal molecular clouds, thus giving us insight into the conditions necessary for star formation.  

Over the last decade, much progress has been made toward systematically characterizing the properties of embedded clusters locally (d $<$ 2.5 kpc) in our own Galaxy (e.g. \citet{ladalada}).  However, what has been missing from these studies is a characterization of embedded clusters in vastly differing physical environments.  Obtaining this information is crucial in making significant advances in our understanding of the origin of embedded clusters.  

The LMC, located at a distance of 48 kpc, is one of the closest galaxies to the Milky Way and provides a unique and ideal laboratory for embedded cluster studies in a physical environment that is different from our own.  Particularly, the LMC has a much lower metal abundance \citep{dufour}, a higher gas to dust ratio \citep{koornneef} and a stronger ultraviolet field \citep{israel} than the Milky Way.  The LMC also provides a sample of molecular clouds that are all approximately at the same distance.  As a result, we can compare embedded cluster properties throughout the region without the ambiguity of varying or uncertain distances, which is a significant issue in Galactic embedded cluster studies.

Astronomers have been working to catalog star clusters and extended objects in the Magellanic Clouds for over fifty-years.  The majority of these studies use optical images to identify and characterize stellar clusters, and as a result, typically only identify clusters older than $\sim$5 Myr (e.g. \citet{bica08,glatt,baumgardt}, and references therein).  Based on these works, $\sim$4200 star clusters have been identified in the LMC, however this census is still incomplete \citep{glatt}.  Previous studies have missed identifying the youngest, embedded clusters, as they are still surrounded by their natal molecular clouds and require NIR observations in order to be identified.  A thorough NIR census of the star-forming content of the molecular clouds in the LMC has yet to be undertaken, but is needed in order to understand whether star and star cluster formation in giant molecular clouds (GMCs) in the LMC differs from what has been found in the Milky Way.  

We have begun a comprehensive and systematic search for embedded clusters in the LMC using the VISTA Magellanic Clouds (VMC) survey \citep{vmc}.  This survey is the first NIR survey of the LMC with resolution and depth suitable for the identification and characterization of embedded clusters.  In combination with VMC data, recently completed high-resolution molecular cloud surveys of the LMC (i.e.\citet{magma}) make it possible to expand our studies of embedded clusters beyond the Milky Way.  We are using these data to identify the young clusters embedded within molecular clouds in the LMC, to measure their basic properties, and finally compare them to the Milky Way embedded cluster sample.  In this paper we present the initial results of our survey for embedded stellar clusters within molecular clouds in the LMC using NIR data from the VMC survey.

\begin{figure*}
\center
\includegraphics[scale=0.9]{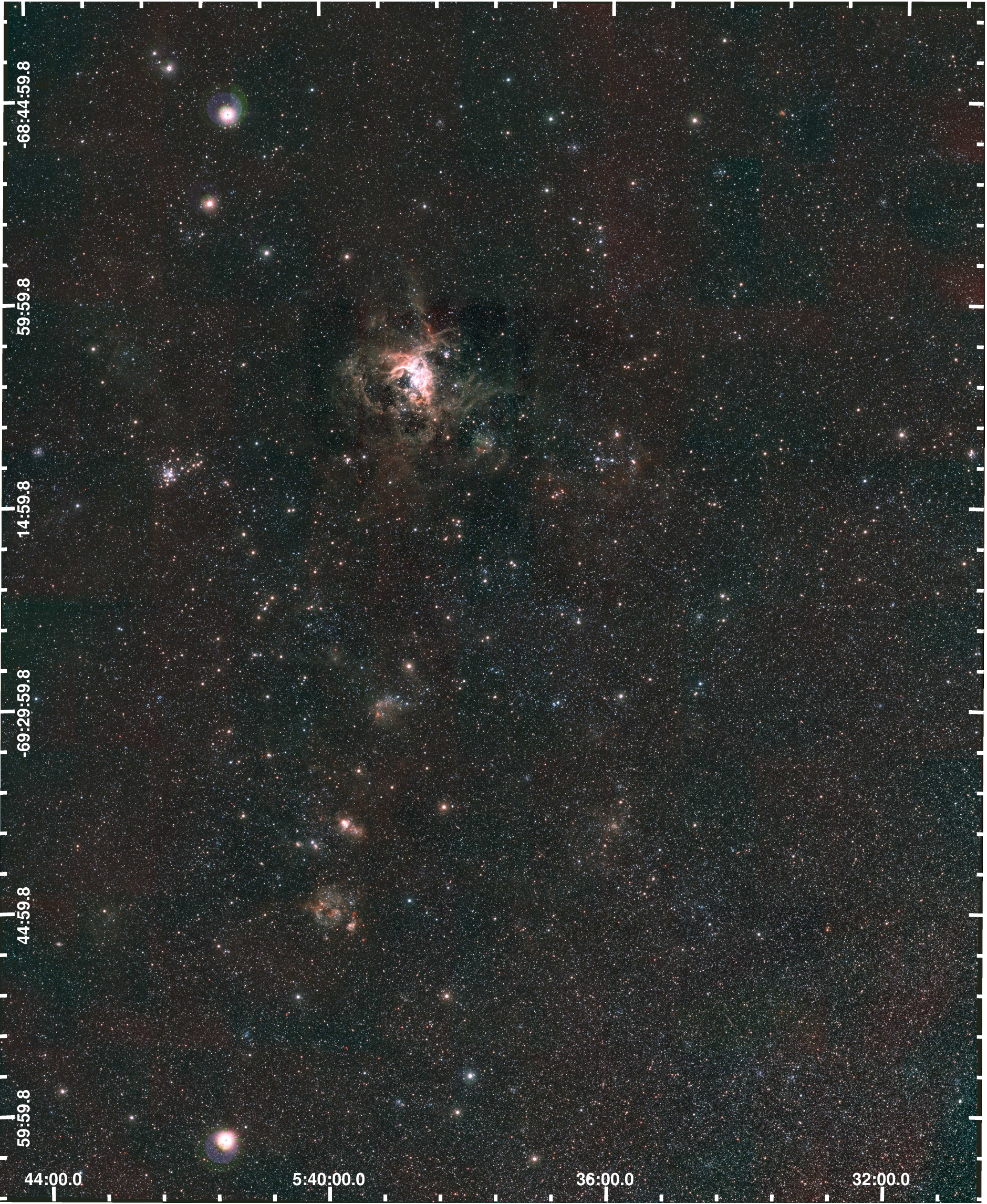}
\caption{A VMC 3-color image (K$_s$ in red, J in green, and Y in blue) of the LMC 6\_6 tile studied in this paper.}
\label{pretty}
\end{figure*}

\begin{figure*}
\center
\includegraphics[scale=0.9]{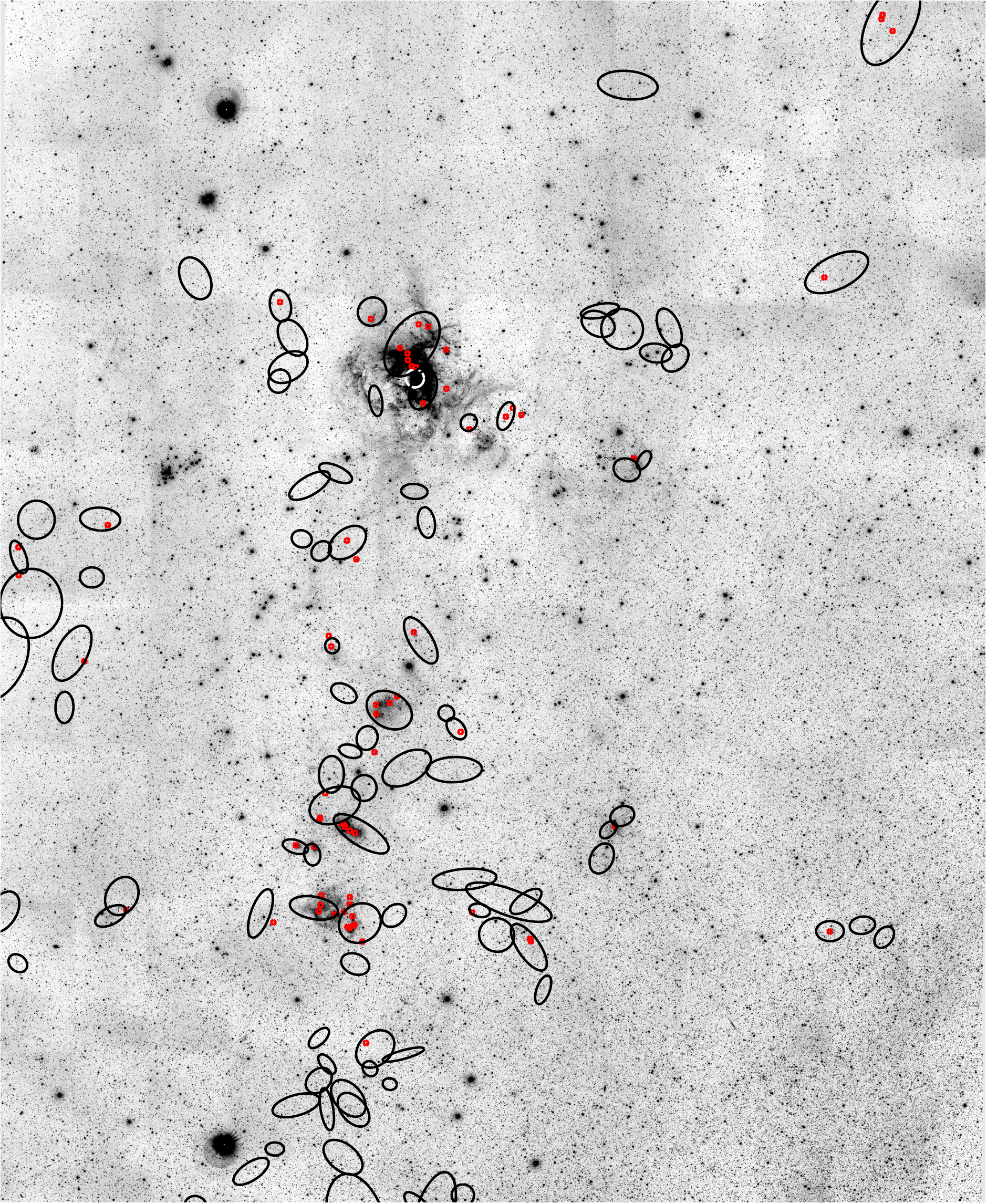}
\caption{A K$_s$ band image of the LMC 6\_6 tile studied in this paper.  The sizes and orientations of the \citet{magma} CO clouds are outlined in black, the boundary of the 40$\arcsec$ radius of the R136 cluster is indicated with a white circle, and the locations of the embedded clusters are marked in red.}
\label{tile}
\end{figure*}

\begin{figure*}
\center
\includegraphics[width=0.9 \textwidth]{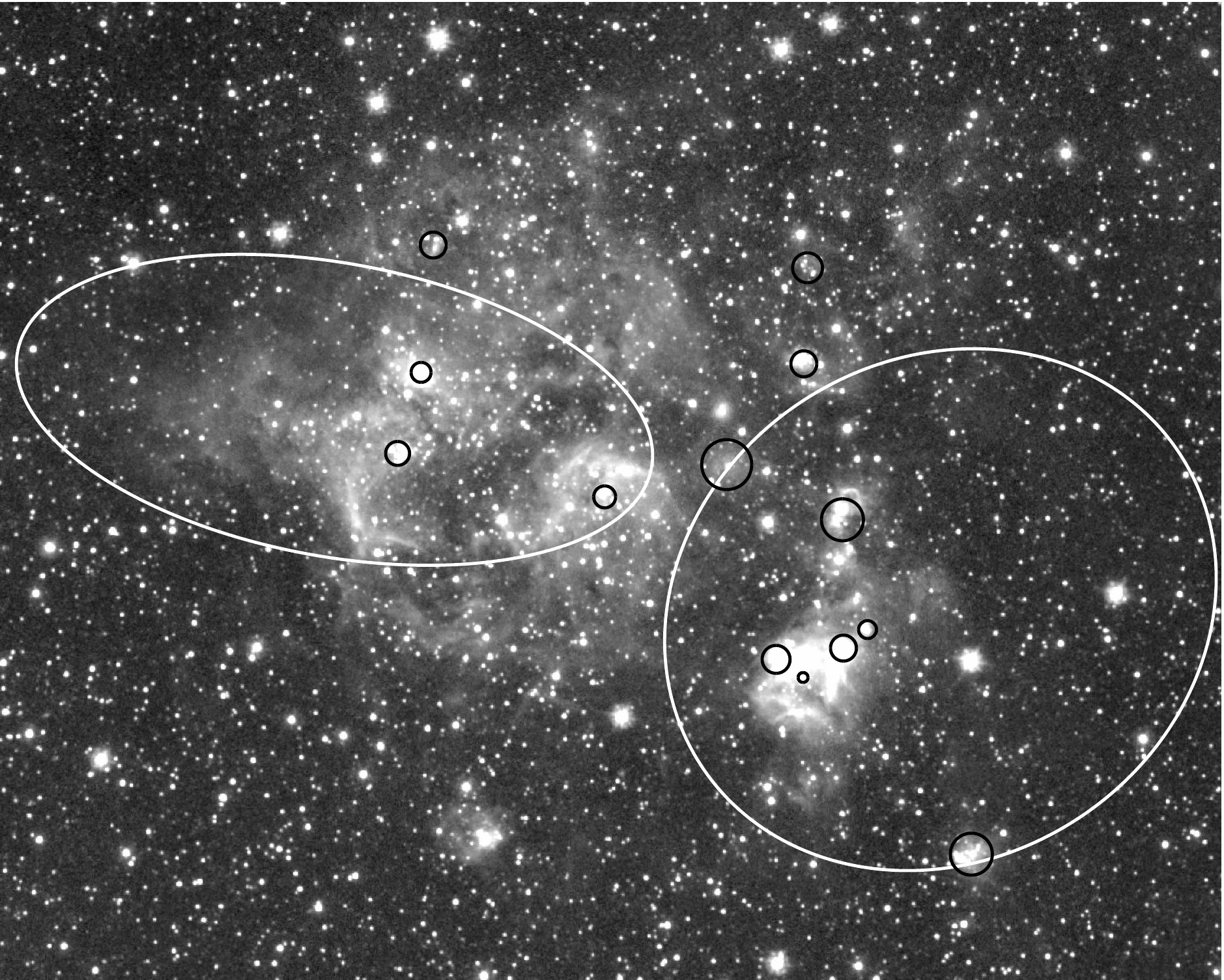}
\caption{Zoomed in images of a sample of the candidate embedded clusters shown in the VMC K$_s$ band.  CO cloud boundaries are in white and the locations of the embedded clusters are marked in black.}
\label{zoom}
\end{figure*}

\section{Data}
We use K$_s$ band images from the VMC survey \citep{vmc} to identify candidate embedded clusters and determine their properties.  The VMC survey is one of the six Visible and Infrared Survey Telescope for Astronomy (VISTA) public surveys and is 70\% complete overall, with 55\% of LMC observations completed  as of October 1, 2015.  The VMC survey is a uniform and homogeneous survey of the Magellanic system in the NIR with VISTA, which is a 4 meter class telescope that was developed in the United Kingdom and is located 1.5 km from the Very Large Telescope (VLT) site in Chile.  The VISTA infrared camera (VIRCAM), which is being used to take the data for this survey, is equipped with a 4x4  array of 16 Raytheon detectors with a mean pixel size of 0.339\arcsec and a field of view of 1.65 deg$^2$.  VIRCAM has a set of broad-band filters: Z, Y, J, H and K$_s$ and a narrow-band filter at 1.18$\mu$m.  Additional details about the telescope and its camera can be found in \citet{emerson} and \citet{dalton}.  VISTA is the largest wide-field NIR imaging telescope in the world and is designed to perform survey observations.  

The VMC survey observes a continuous area of sky, filling in the gaps between the detectors, by observing a sequence of six {\it pawprint} positions for each region, each offset by a significant fraction of the detector.  Individual pawprints  cover an area of 0.59 deg$^2$.   A  VMC tile is produced by combining 96 different images (16 individual detector images per each of the 6 pawprints) and covers  an area of $\sim$1.65 deg$^2$.  Each region of the sky contained within the resulting VMC tile is therefore observed at least twice, except for two edge strips in the extreme "Y" directions of the array.  

The VMC survey is a NIR Y, J, and K$_s$ survey of the Magellanic system.  The entire LMC will be covered by 68 tiles, which are identified by two numbers that indicate their position in the mosaic that covers the entire system.  Observations are obtained in service mode by ESO staff.  This guarantees efficiency of operations and a high level of homogeneity.  Data reduction is completed by the VISTA Data Flow System (VDFS) pipeline at the Cambridge Astronomical Survey Unit (CASU) and at the Wide Field Astronomy Unit (WFAU).  This pipeline is specifically designed for reducing VISTA data \citep{irwin} and is used to process up to 250 GB per night of data.  

In this paper, we use K$_s$ band data from the LMC 6\_6 tile of the VMC survey. The data were reduced onto the VISTA photometric system (Vegamag = 0) with the VDFS pipeline v1.1 and extracted from the VISTA Science Archive (VSA) \citep{cross12} using data release VMC DR1. The total exposure time for the LMC 6\_6 image is 9372 seconds, which corresponds to a limiting magnitude of 21.5 mag at 5$\sigma$, which is considerably more sensitive than the existing 2MASS survey \citep[K $<$ 14][]{2MASS} .  A 3-color image of the LMC 6\_6 tile is shown in Figure~\ref{pretty}.  Additionally, the resolution is considerably better than  2MASS (2$\arcsec$), with the average FWHM of the LMC 6\_6 tile being $\sim0.92\arcsec$.    Additional information about the VMC survey data and initial results can be found in \citet{vmc}.

\begin{figure}
\includegraphics[width=0.5 \textwidth]{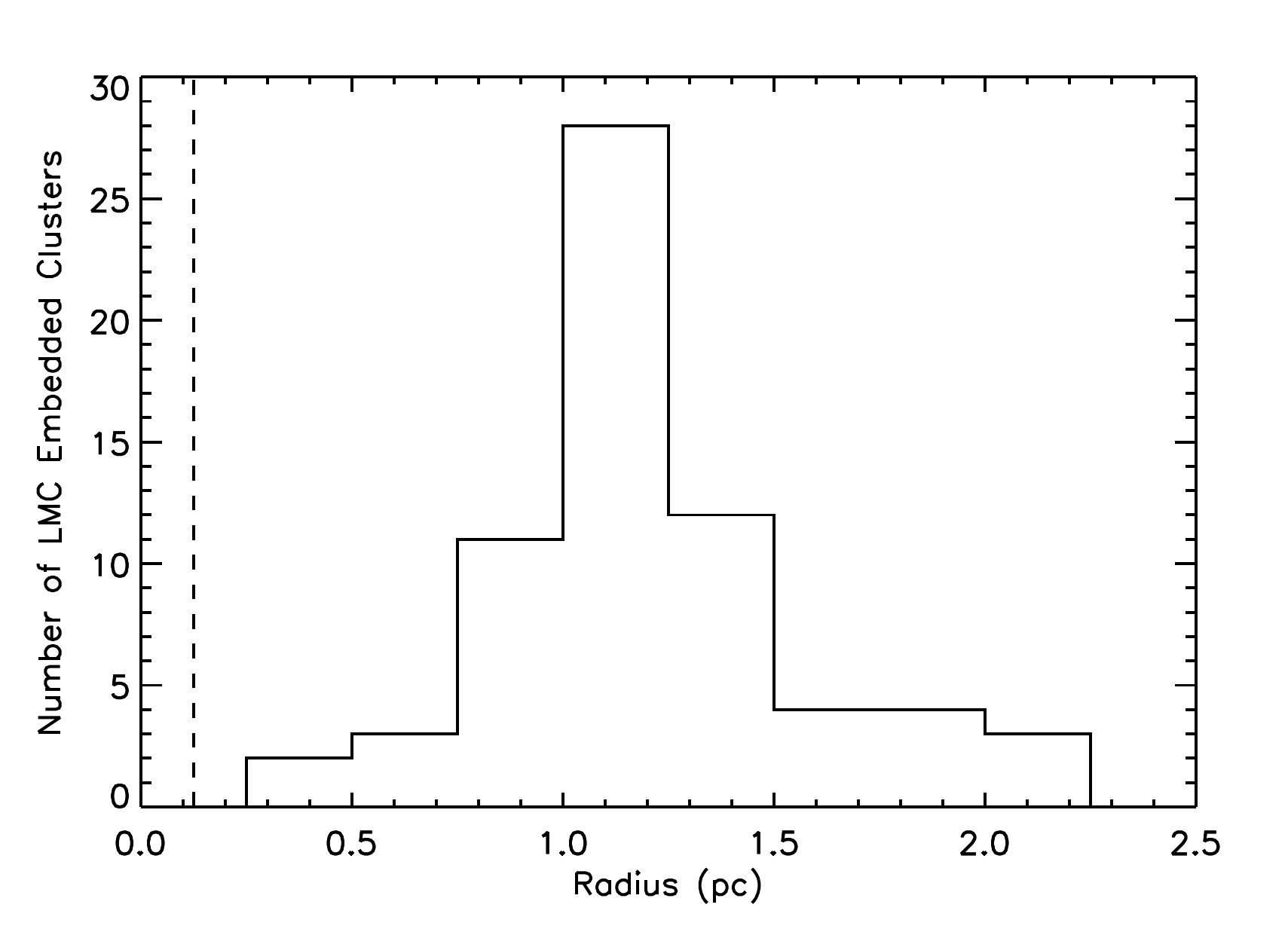}
\caption{This Figure shows the size distribution of our 67 embedded cluster candidates with the size in pc on the x-axis and the number of clusters along the y-axis with a bin size of 0.25 pc.  The value of the VMC data's seeing limit is represented by the dashed line.}
\label{sizedist}
\end{figure}

\section{Results}

\subsection{Identification of Clusters}
At the distance of the LMC, 1\arcsec corresponds to $\sim$0.24 pc.  Galactic embedded clusters are typically $\sim$1 pc in diameter \citep{ladalada}, which, given the $\sim1\arcsec$ resolution of the VMC data, indicates that we should be able to resolve the embedded clusters in the VMC data.  While we are not able to resolve all of the embedded cluster members, we are able to resolve some that are away from the cluster centers where crowding is less of an issue.

Cluster candidates were identified via visual inspection.  As we are interested in the youngest clusters, we used the K$_s$ images for our identification to minimize the effects of extinction.  We first looked for density enhancements above the general stellar field and/or irregularly shaped extended sources that were suggestive of an unresolved cluster center.  We examined the images a total of 3 times, independently searching for clusters each time.  In order to be considered an embedded cluster candidate, a cluster had to (1) have an enhanced density of stars above the general background and/or have an extended component that was suggestive of an unresolved core or nebulosity.  In the cases where the cluster appeared mostly as an extended/nebulous component, it was also required that several point sources to be present in order to distinguish the cluster candidate from clumps of nebulosity.  Next, to ensure the youth and embedded nature of the clusters, we required that the candidates (2) be associated with molecular (CO) gas.  The purpose of criterion 1 is to identify candidate clusters.  Criteria 2 is used to ensure the youth and embedded nature of our candidates.

Using the results of the MAGMA CO survey, we were able to obtain the size and orientations of the 77 MAGMA CO clouds in the LMC 6\_6 tile.  The locations and sizes of these molecular clouds are outlined in black in Figure 2. Cluster candidates located either within the boundary of a molecular cloud or within $\sim$3 cluster diameters of a cloud boundary were considered to be associated with that cloud.  When clouds overlapped, or when clusters were located between cloud boundaries, we consulted the \citet{magma} CO contours in order to determine which cloud was most likely to contain the cluster based on proximity.  This was only necessary for 8 clusters.  

Using these criteria, we were able to identify 75 embedded cluster candidates,  associated with CO clouds.  A number of these candidates were found close to the R136 cluster and may be part of this super star cluster.  Indeed, \citet{R136} found that the luminosity profile of R136 was continuous out to 40$\arcsec$ from the cluster center.  Therefore, we decided to exclude the area within a 40$\arcsec$ radius of the center of R136 and removed 8 clusters from our original candidate list.  Our final embedded catalog for the LMC 6\_6 tile contains a total of  67 embedded cluster candidates.  Twenty-two of these candidates were previously identified as clusters or stellar associations and 12 were identified as emission nebulae \citep{bica08}.  Therefore 45 of our cluster candidates are newly identified as embedded clusters, and 33 candidates are newly discovered objects.  The locations of the embedded clusters in our catalog are marked in red in Figure~\ref{tile}.  Figure~\ref{tile} illustrates that the embedded cluster candidates are primarily located in the molecular clouds near the 30 Doradus star-forming region and throughout the molecular ridge which extends south from the R136 super star cluster.  This is where the majority of the CO resides in the LMC 6\_6 tile.  A zoomed in image of two molecular clouds and candidate embedded clusters can be found in Figure~\ref{zoom}.

\subsection{Cluster Properties}
Determining even the basic properties of embedded clusters is a vital step in constructing a complete picture of their formation and evolution.  Therefore, the first step in our analysis was to determine cluster sizes, integrated luminosities and masses.  We then used these basic properties and compared them to the properties of their natal molecular clouds.  

\subsubsection{Sizes}
\label{sizes}

We determined the sizes of the cluster candidates using a luminosity profile analysis.  The average FWHM of the VMC data is $\sim$0.92$\arcsec$, which corresponds to a radius of approximately 1.5 pixels.  Therefore, following the assumption that clusters would be larger than the stellar FWHM, we measured the integrated luminosity of each cluster using apertures of increasing radii from 2 to 43 pixels, in steps of 1 pixel.  The apertures were centered on the pixel having the peak luminosity.   We then plotted the integrated luminosity as a function of aperture size. The radius of the cluster was defined as the point at which the luminosity profile turns over and the increase in brightness as a function of aperture radius levels off. Operationally, this was chosen to be equal to the radius of the first aperture for which the average of the integrated magnitudes of the adjacent four points on the luminosity profile has a standard deviation that is less than or equal to 0.1 magnitudes.  Four of the embedded cluster candidates did not have radii that were well defined by this method.  In these cases, we determined their sizes using the aperture at which the slope of the luminosity profile was equal to $-0.1$ magnitude/pixel and verified the size via visual inspection.  

The size distribution of our embedded clusters can be seen in Figure~\ref{sizedist} with cluster radius in parsecs on the x-axis and the number of clusters on the y-axis.  The distribution spans a range of $\sim0.25$-2.25 pc.  There is a peak at $\sim1.1$ pc then the distribution falls off sharply as the sizes increase. 

\begin{figure}
\includegraphics[width=0.5 \textwidth]{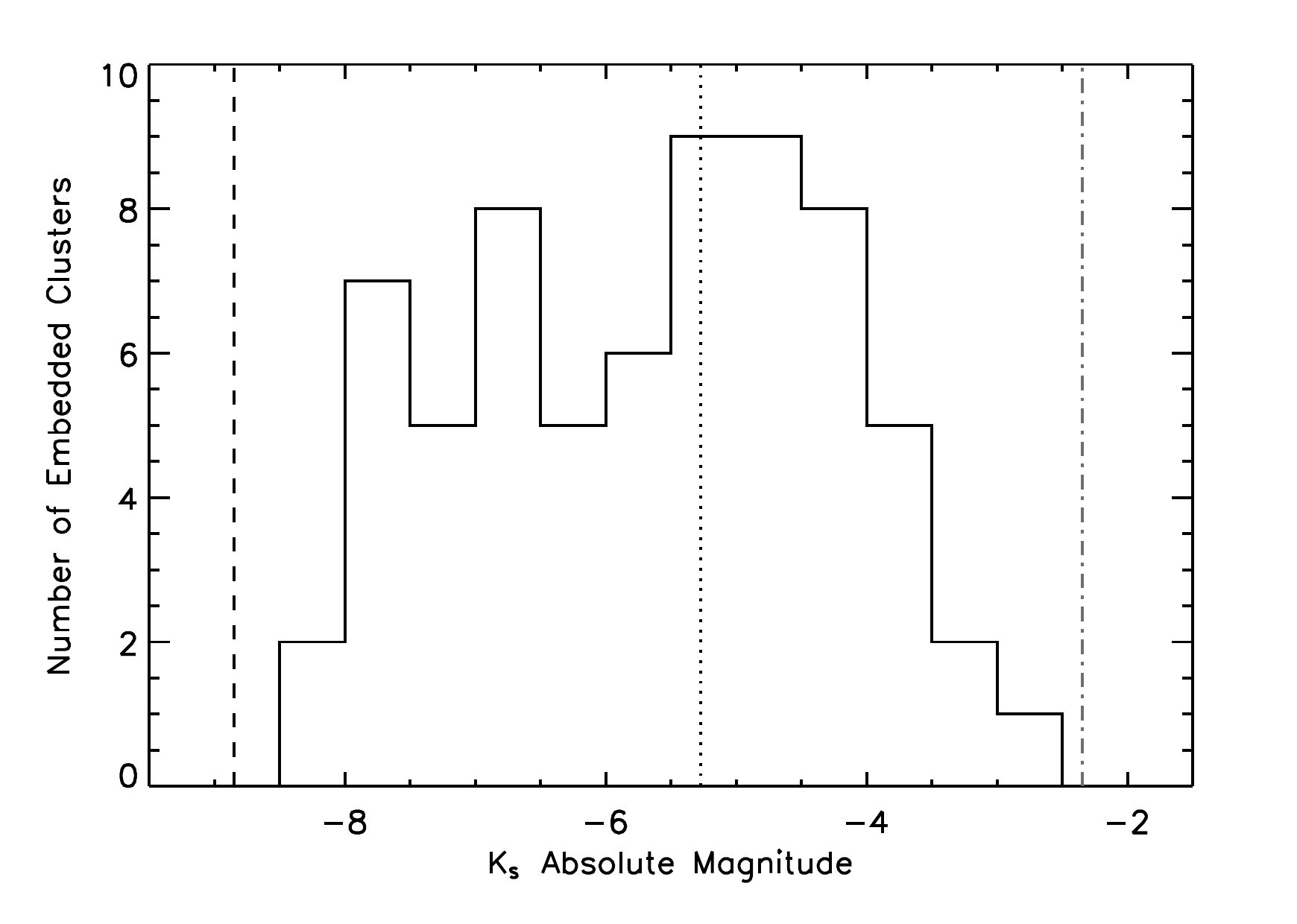} 
\caption{This shows the K$_s$ band luminosity function for the clusters in our embedded cluster catalog with absolute K$_s$ magnitude plotted on the x-axis and the number of embedded clusters plotted on the y-axis and bins that are 0.5 magnitude in width.  The absolute magnitudes of the Trapezium cluster and massive Milky Way cluster NGC 3603 are shown for reference by the black vertical dotted and dashed lines respectively.  The roughly determined cluster detection limit of the VMC data is plotted in the gray dash-dotted line.}
\label{LF}
\end{figure}

\subsubsection{Luminosities}
\label{luminosities}

We measured the integrated cluster magnitudes for each cluster in the sample using DAOPHOT with apertures that were defined by the individual cluster sizes determined in Section~\ref{sizes}.  The sky annuli that were used for all of the clusters had an inner sky radius of 43 pixels and an outer sky radius of 48 pixels.  We investigated the effect of the choice of sky annulus by measuring integrated cluster luminosities by using sky annuli ranging from 21 pixels to 48 pixels.  We found that the change in luminosity due to changes in the sky annulus was never more than 0.13 magnitudes.  The magnitudes of the clusters were calibrated using the zeropoints and exposure times in the image headers via the methods described in the CASU VIRCAM documentation (http://casu.ast.cam.ac.uk/surveys-projects/vista/\\technical/catalogue-generation) and using a distance modulus of 18.5.  The K$_s$ band Luminosity Function (KLF) of our embedded clusters is shown in Figure~\ref{LF}, with K$_s$ band absolute magnitude on the x-axis and the number of clusters on the y-axis.  The number of embedded clusters generally increases with increasing magnitudes, with a peak at approximately -5 mag and then there is a drop off in the numbers of clusters.  This drop off is likely due to incompleteness at the faint end of the KLF.

The integrated magnitudes of the Trapezium and NGC 3603 embedded clusters shown in Figure~\ref{LF} were determined from de-resolved K band 2MASS images using the same method for size determinations as discussed in Section~\ref{sizes}.  We then used DAOPHOT in the same way as detailed above to get integrated cluster luminosities.  These cluster magnitudes were calibrated using the zeropoints in the image headers as described in the 2MASS documentation (http://www.ipac.caltech.edu/2mass/releases/allsky/faq.\\html\#conversion).  The dotted line in Figure~\ref{LF} indicates our estimated cluster detection limit for the LMC embedded cluster detections.  We calculated this limit by placing 500 1 pc apertures in areas of the VMC LMC 6\_6 image that did not contain any stars or nebulosity, then used DAOPHOT to get the flux in each of the background apertures.  DAOPHOT returned fluxes for 381 of the 500 apertures, the rest contained errors due to the apertures containing pixels below the low-bad pixel threshold in DAOPHOT.  We used the value of the flux that corresponded with the peak of the distribution of the 381 valid background apertures as a baseline value of the noise in our data.  We then multiplied the noise counts by ten in order to calculate the minimum signal with a S/N=10 which we could detect in a 1 pc radius aperture in our data.  The result is the estimate of the cluster detection limit (M$_{K_s}\sim-2.35$).
 
The context of the luminosities of the Trapezium and NGC 3603 helps to reveal that the LMC contains more luminous clusters than the local environment.  The Trapezium, which is the most luminous embedded cluster in the solar neighborhood, is about average in its luminosity when compared to the LMC embedded cluster sample.  In fact, the Trapezium falls in the middle of the LMC cluster luminosity range, and roughly at the peak of the LMC luminosity function.  This suggests that clusters similar to the Trapezium are relatively common in the LMC embedded cluster sample.  While a sizable fraction of  the LMC embedded clusters are more luminous than the Trapezium, they are all fainter than NGC 3603, which is located in the Carina spiral arm and is one of the most luminous young clusters in the Milky Way.  We should note however, that R136 in 30 Doradus (M$_{K_s}\sim-12$) is considerably more luminous than NGC 3603.  Finally, based on our rough cluster detection limit calculation, we find that we should be able to detect embedded clusters in the LMC that are fainter than the Trapezium.  However, while we may be sensitive to these lower luminosity clusters, we identify few of them in our LMC data.  This is most likely due to completeness issues or biases in our cluster identification techniques.

\begin{figure}
\includegraphics[width=0.5 \textwidth]{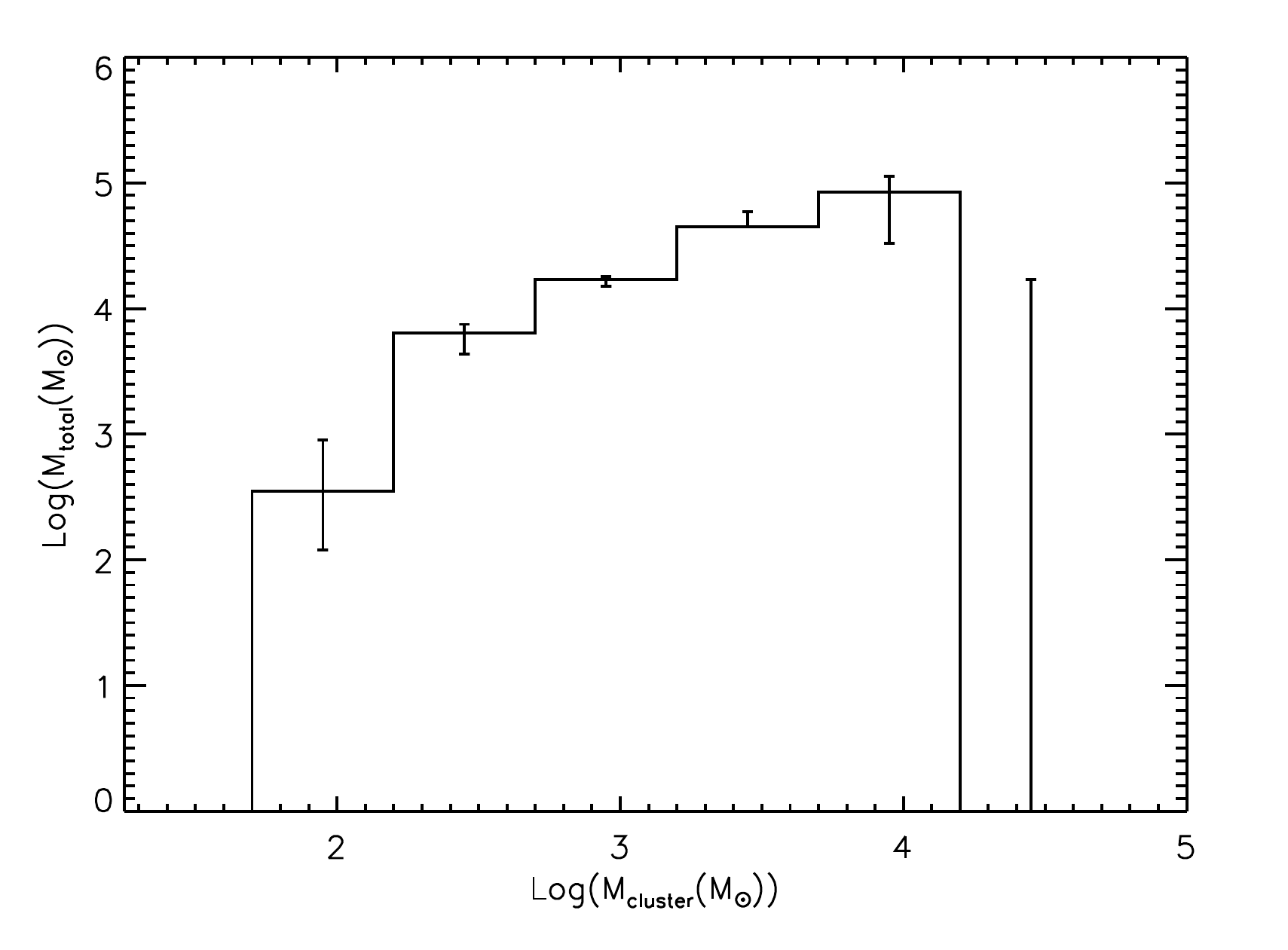}
\caption{This shows the embedded cluster mass function (ECMF) for the LMC 6\_6 tile.  This plot displays the distribution of the log of the total cluster mass as a function of the log mass.  The boundaries of the logarithmic bins of this plot were chosen to match the bins used in \citet{ladalada} for ease of comparison between the Milky Way ECMF and the LMC ECMF.  The error bars are indicative of the maximum differences of the height of the bins if one changes the age assumption in the mass estimation models to 1, 2, 3, 5 or 10 Myrs. } 
\label{massfxn}
\end{figure}

\subsubsection{Masses} 
\label{masses} 

In order to determine mass estimates for our embedded cluster candidates we use a monte carlo model to create simulated clusters with masses from 100M$\sun$ to 10$^5$M$\sun$ at intervals of 10M$\sun$ until 1000M$\sun$, above which we create clusters at intervals of 100M$\sun$.  The model clusters were constructed assuming a Trapezium IMF \citep{muench} with a minimum stellar mass of 0.1 M$\sun$ and a maximum stellar mass of 18 M$\sun$, keeping track of each star's mass until the simulated cluster reaches the desired total mass.  Then, we determine the VISTA K$_s$ luminosity for each star in the simulated cluster using PARSEC stellar evolution models (\citealt{bressan}, \citealt{tang}, \citealt{chen}) with an assumed age and LMC metallicity (Z=0.008).  An age of 2 Myrs was selected in order to be consistent with age assumptions in \citet{ladalada} and for ease of comparisons to their results, however we determined embedded cluster masses for a range of ages from 1-10 Myrs in order to examine the effects of age on our results.  The integrated cluster luminosity is then calculated by summing up the individual stellar luminosities in the cluster.  Simulated clusters were created 100 times for each cluster mass and the variations in the resulting integrated luminosities for each cluster mass were used to calculate the errors in our mass estimates.  We used the calculated luminosities for each simulated cluster mass to create a mass-luminosity model grid, from which we interpolated our embedded cluster masses from their integrated luminosities.  Our initial calculations indicate that the mass estimates of the embedded clusters have errors that are $\leq$20\%.   However, one must also keep in mind that these mass estimates are based on integrated luminosities alone, and therefore likely include a non-stellar component that would artificially increase the luminosity of the embedded clusters.  The contribution from non-stellar sources within the cluster cannot be separated from the unresolved stellar component of these LMC embedded clusters and the amount of luminosity added is variable from cluster to cluster.  A comparison between the Trapezium luminosity as determined from integrating it's KLF and its luminosity as determined in Section~\ref{luminosities} finds that the integrated luminosity of the Trapezium is 4.14 times greater than the luminosity of the stars alone.  This equates to a factor of 3.77 difference in the mass when using our mass estimation method.  For an additional comparison point, when we do the same comparison for the Milky Way cluster IC348, there is only an 8$\%$ difference in the masses between using the integrated luminosity and the luminosity of only the stars in the cluster.  Therefore, our mass estimations are likely accurate within a factor of 4, but should be considered upper limits.  A detailed discussion of the models and error estimates will be presented in a forthcoming paper.  

The derived cluster masses in our catalog range from $\sim$100 M$\sun$ to $\sim1.4 \times 10^4$.  Figure~\ref{massfxn} shows the Embedded Cluster Mass Function (ECMF) for our LMC embedded cluster catalog.  The method for plotting this ECMF is the same as the one used in \citet{ladalada} for ease of comparison with their Milky Way embedded clusters.  The ECMF was derived by summing individual embedded cluster masses (M$_{ec}$) in evenly spaced logarithmic bins, beginning with Log(M$_{ec}) =$1.7.  The width of the logarithmic bins helps to absorb the uncertainty in the mass estimates.  The ECMF is equal to M$_{ec} \times \frac{dN}{dlog(M_{ec})}$ and therefore differs from the mass function ($\frac{dN}{dlog(M_{ec})}$) of embedded clusters by a factor of M$_{ec}$.  The ECMF increases with increasing cluster mass, which would indicate that the majority of the total mass of the LMC embedded clusters resides in the more massive clusters.  However, our survey likely suffers from incompleteness at the low mass end of the ECMF.  If we estimate the completeness of our mass estimates based on the location of the turn-over of the LF (Figure~\ref{LF}), below which we are likely incomplete, the ECMF is incomplete for masses below $\sim$600 M$_{\sun}$.  This incompleteness could be driving the shape of the LMC ECMF, as increasing our completeness at the low mass end of the distribution, would flatten the ECMF.   

\begin{figure}
\includegraphics[width=0.5 \textwidth]{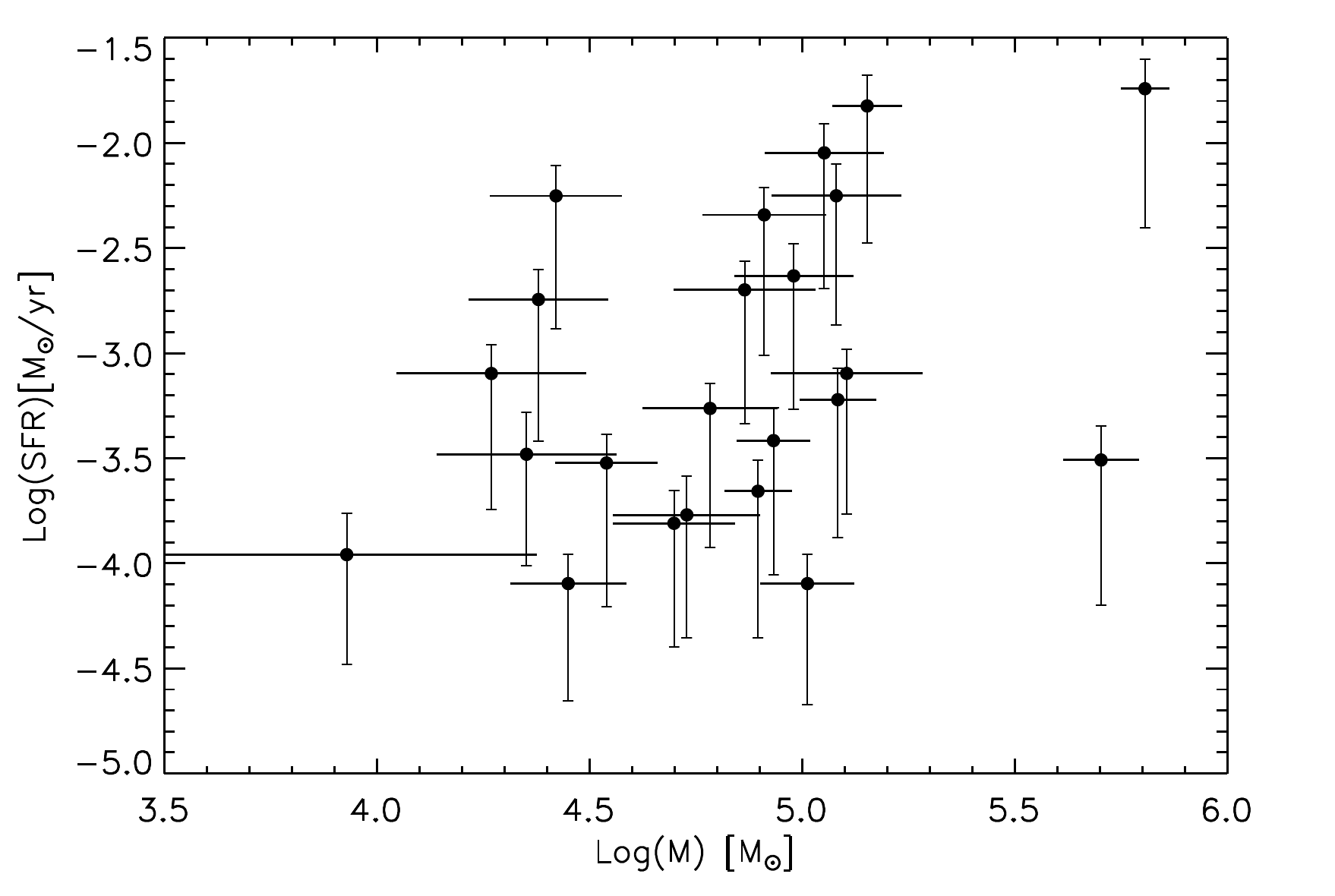} 
\caption{This shows the SFR versus molecular cloud mass for the clusters in our embedded cluster catalog.  The vertical error bars represent the changes in the SFR if one changes the age assumption in our calculations to 1 and 10 Myrs.  The horizontal error bars correspond to the errors in the molecular cloud masses as reported in \citet{magma}.}
\label{sfr}
\end{figure}

\section{Star Formation Rates}

Using the mass estimates for our embedded cluster sample, we were able to calculate Star Formation Rates (SFRs) for each of the molecular clouds in the LMC 6\_6 tile that contain embedded clusters.  For our calculations we assumed a cluster age of 2 Myrs, which is consistent with the assumed cluster age that we used for our mass calculations.  Figure~\ref{sfr} shows the SFRs for the molecular clouds in our sample with the log of the molecular cloud mass plotted along the x-axis and the log of the SFR plotted along the y-axis.  There does not appear to be a particular trend between the CO mass of the molecular clouds and their corresponding SFRs as a wide range of SFRs are possible at the full range of molecular cloud masses.  In fact, when a line is fit to the data, the fit has a correlation coefficient of 0.404, indicating that there is no trend.  While some of the variations in the SFRs are likely due to uncertainties in our cluster mass estimations, the remainder is likely real and may be due to differing physical conditions in the molecular clouds in our sample.  

\section{Discussion}
\label{discussion}

Previous studies of embedded clusters have focused on the local environment, particularly within $\sim$2 kpc of the Sun (e.g. \citet{ladalada}).  While observations over the past 25 years have established the importance of embedded clusters in understanding the star-formation process, our knowledge of the fundamental properties of these clusters remains incomplete, especially beyond the local environment. This paper presents the initial results of the first large scale census of embedded clusters within GMCs in the LMC, a significantly different environment from the local Milky Way.

\subsection{Comparison to the Milky Way}

\begin{figure}\includegraphics[width=0.5 \textwidth]{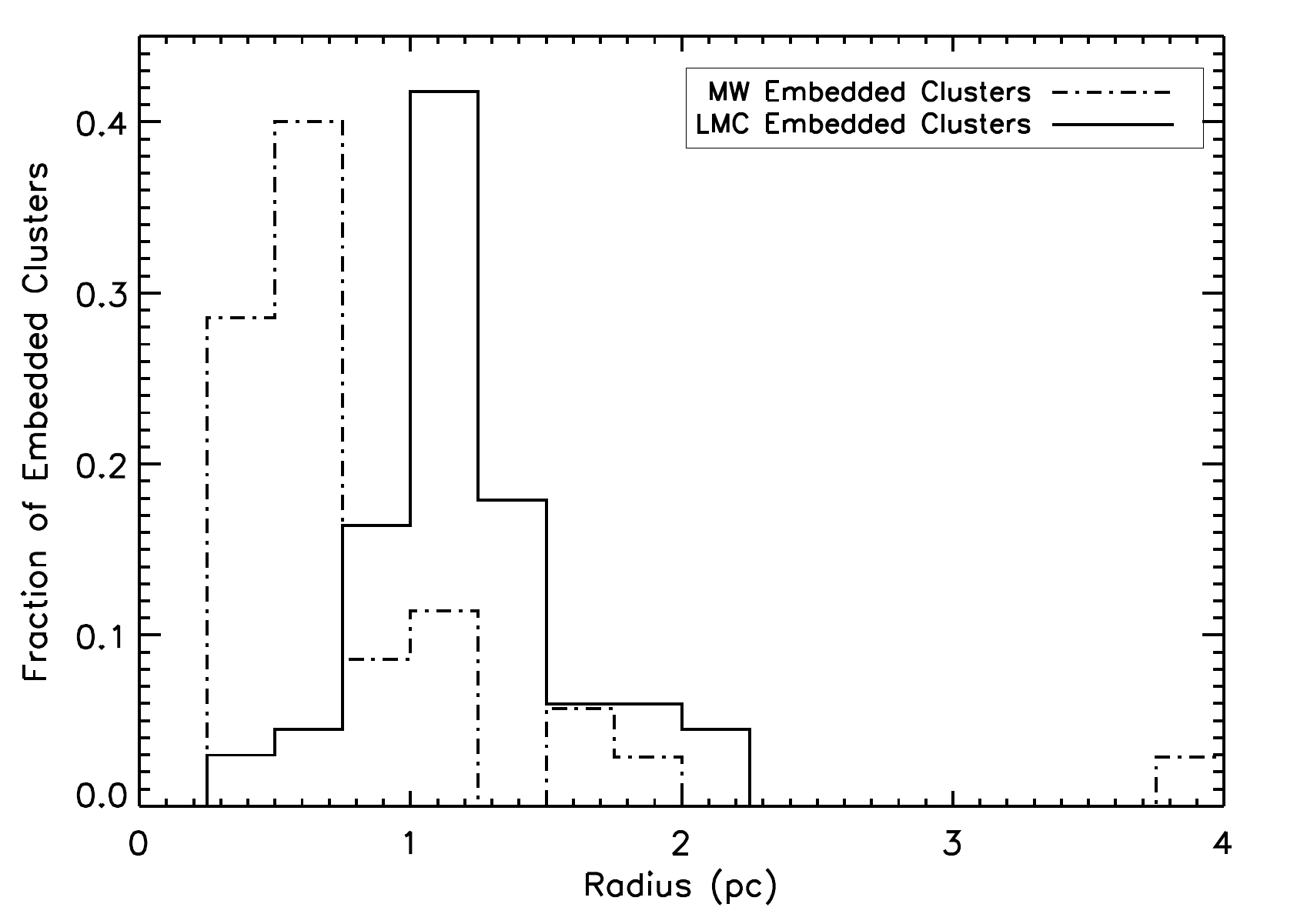}
\caption{This shows the normalized size distributions of the LMC embedded clusters in the solid line and Milky Way embedded clusters in the dot-dashed line.  The Milky Way embedded cluster sizes come from \citet{ladalada}.}
\label{size_compare}
\end{figure}

\subsubsection{Sizes}
\label{sizes2}
The normalized size distributions of the LMC 6\_6 tile embedded clusters and the Milky Way embedded clusters from \citet{ladalada} are shown in Figure~\ref{size_compare}.  The range in sizes for both distributions are similar, with the exception of one Milky Way embedded cluster - the Trapezium.  Additionally, the overall shapes of the size distributions in both galaxies are consistent, with both having a pronounced peak with a tail that goes to larger sizes.  It is possible that the LMC embedded clusters are, in fact, larger than the Milky Way embedded clusters, however, since the clusters sizes are determined in different ways and based off of data of different depths, it is difficult to say this for certain.  In a forthcoming paper, we will address this uncertainty by completing a suite of artificial cluster tests in order to directly compare the actual sizes of clusters to the sizes measured using our method.  This will allow us to more accurately determine the errors in LMC embedded cluster size distribution, which will assist in our comparison to the Milky Way embedded cluster population.

\subsubsection{Surface Densities}
\label{surfacedensities}
Embedded clusters appear to be common in the LMC.  Our search for embedded clusters covers only a small portion of the LMC as a whole, and already we have identified approximately the same number of embedded clusters in the LMC as have been identified within $\sim$2 kpc of the Sun \citep{ladalada}.  This is consistent with the observation that the LMC has a large general cluster population of several thousand clusters and has ongoing star formation.  In order to quantify the commonality of embedded clusters, we determined the surface density of embedded clusters in the LMC 6\_6 tile and compared it to the surface density of embedded clusters within 500 pc of the Sun where the Milky Way embedded cluster survey is most complete.  Taking into the account the 30.7$^{\circ}$ inclination of the LMC \citep{nikolaev} in our calculation of the surface area, we found that the physical area covered by the LMC 6\_6 tile is $\sim$1.4 kpc$^2$.  Therefore, the LMC embedded cluster surface density is $\sim$48 clusters/kpc$^2$.  The surface density of embedded clusters in the Milky Way within 500 pc of the Sun above our estimated LMC cluster detection limit (see Figure~\ref{LF}) is $\sim$14 clusters/kpc$^2$.  This reveals that embedded clusters are $\sim$3.4 times more common in the LMC 6\_6 tile than in the local environment.  

In order to further compare the local Milky Way embedded cluster population with our LMC embedded clusters, we calculated embedded cluster mass surface densities for each catalog.  The total embedded cluster mass for the LMC 6\_6 tile is $\sim1.53 \times 10^5$ M$_{\sun}$, dividing by the area in the tile, the mass surface density for our LMC embedded clusters is $\sim$0.1 M$_{\sun}$/pc$^2$.  For the Milky Way within 500 pc of the Sun, the total embedded cluster mass for clusters above our estimated LMC cluster detection limit is $\sim2.0 \times 10^3$ M$_{\sun}$, which makes the local Milky Way embedded cluster mass surface density 0.0025 M$_{\sun}$/pc$^2$.  The LMC embedded cluster mass surface density is about 40 times greater than the local embedded cluster mass surface density.  

When we use a similar method to compare the molecular cloud mass surface density between the two environments we find that the molecular cloud mass surface density of the LMC 6\_6 tile is 2.77 M$_\sun/$pc$^2$.  This is $\sim$3.2 times greater than the molecular cloud surface density of the Milky Way within 500 pc of the Sun, which is 0.87 M$_\sun/$pc$^2$ \citep{dame2001}.  When we combine the embedded cluster and molecular cloud mass surface densities, we calculate a embedded cluster formation efficiency for each environment.  The LMC 6\_6 tile efficiency is 3.6\%, whereas the embedded cluster efficiency for the Milky Way within 500 pc of the Sun is $\sim$0.3\%.  Therefore, the LMC seems to be about an order of magnitude more efficient when forming embedded clusters.  

It appears that the local Milky Way and the LMC 6\_6 tile have a similar frequency of embedded clusters when we compare the number of clusters to the molecular cloud mass density, as both calculations are about a factor of 3 higher in the LMC than in the local Milky Way.  Despite this, the embedded cluster mass surface densities differ by about 50 times.  However, we must keep in mind that the embedded cluster masses are to be treated as upper limits.  In Section~\ref{masses}, we estimate that our cluster mass estimates are accurate within a factor of 4.  If we take this into account when considering the embedded cluster mass surface density, the surface density would still be an order of magnitude higher in the LMC.  This suggests that the LMC molecular clouds are forming more massive embedded clusters than the molecular clouds within 500 pc of the Sun, which is also supported by the increased embedded cluster efficiencies in the LMC.  The elevated efficiency and embedded cluster masses that we see in the LMC may be due to the particular environment in the LMC 6\_6 tile.  It seems that this is a region with stimulated star formation, and may not be representative of the LMC in general.  It is also possible that the lower metallicity of the LMC is affecting the star formation process and allowing for the formation of more massive clusters.

\begin{figure}
\includegraphics[width=0.5 \textwidth]{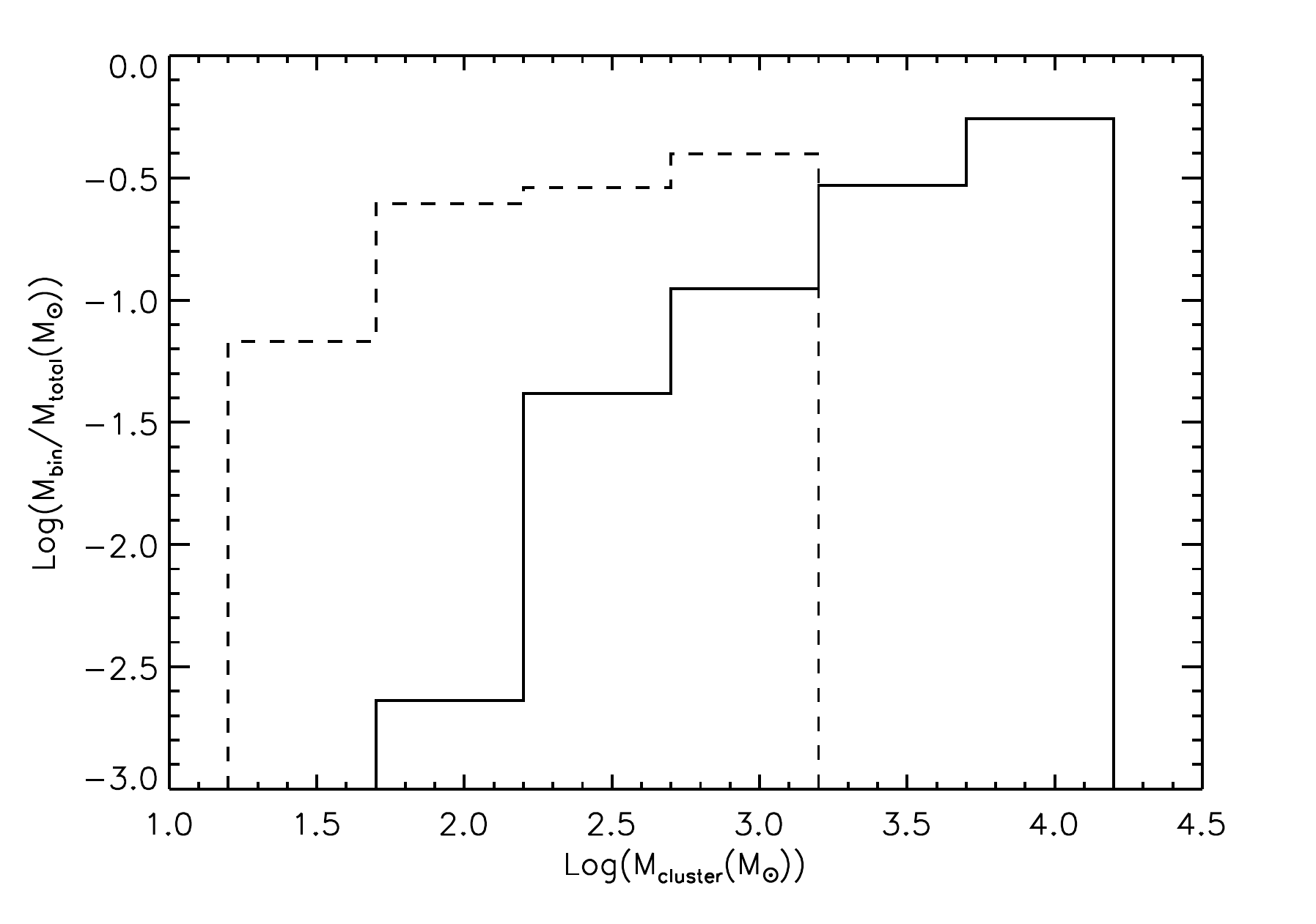} 
\caption{This figure shows the LMC embedded cluster mass function in the solid line and the \citet{ladalada} embedded cluster mass function in the dashed line.  These functions were derived by summing individual embedded cluster masses (M$_{ec}$) in evenly spaced logarithmic mass bins, 0.5 dex in width, beginning at Log(M$_{ec}$) = 1.2.  Additionally, both functions were normalized by dividing each bin by the sum of the embedded cluster masses of each distribution in order to more easily compare the shape of the mass functions.}
\label{massfxn_compare}
\end{figure}

\subsubsection{Embedded Cluster Mass Function}

The embedded cluster population in the LMC 6\_6 tile is generally more massive than the embedded cluster sample from \citet{ladalada}.  The local Milky Way embedded cluster masses span a range from 20-1100 M$_{\sun}$, whereas our LMC embedded clusters range from 100 M$_{\sun}$ to 14000 M$_{\sun}$.  The highest mass clusters in each sample differ in mass by more than an order of magnitude, but the distributions overlap for $\sim$1.5 orders of magnitude.  For ease of comparison between the two ECMFs, we have normalized the distributions, and the results are shown in Figure~\ref{massfxn_compare}.  From this figure it is clear that the LMC embedded clusters are generally more massive than the clusters in the local Milky Way.  One can also see that the overall shapes of the two ECMFs appear to be different.  Overall, the LMC ECMF increases with increasing cluster mass, whereas the Milky Way ECMF is largely flat.  However, it is likely that the LMC ECMF is incomplete below $\sim$ 600 M$_{\sun}$, as discussed in Section~\ref{masses}, and this incompleteness may artificially create a greater slope in the LMC ECMF.  Taking this into account, we consider the overall shapes of the Milky Way and LMC ECMFs together, looking primarily at the highest two bins of the LMC ECMF, and the highest three bins of the MW ECMF, where they are likely the most complete, in order to gain a sense of the cluster formation paradigm over a large range of masses.  Based on this comparison, it is possible that an overall ECMF would be relatively flat from $\sim$50 M$_{\sun}$ to $\sim$14000 M$_{\sun}$.  A flat ECMF would indicate that while high mass clusters are comparatively rare, they contribute a similar amount to the total embedded cluster mass as the more numerous low mass clusters, which is what is seen in the local Milky Way ECMF.  

It is also possible, that despite our incompleteness, the LMC ECMF in fact has a positive slope where the mass function increases with increasing cluster mass.  If this is the case, it would indicate that the majority of the total mass of the LMC embedded clusters resides in the more massive clusters and that the LMC is preferentially forming massive embedded clusters.  This explanation is supported by what we see with the mass surface densities in Section~\ref{surfacedensities}.  This could be an effect of the lower metallicity environment in the LMC.  Expansion of our embedded cluster search to additional regions in the LMC as well as more sensitive and higher resolution observations of low mass clusters are needed in order to conclusively determine whether the LMC ECMF is flat or increases as a function of increasing mass.

\begin{figure}
\includegraphics[width=0.5 \textwidth]{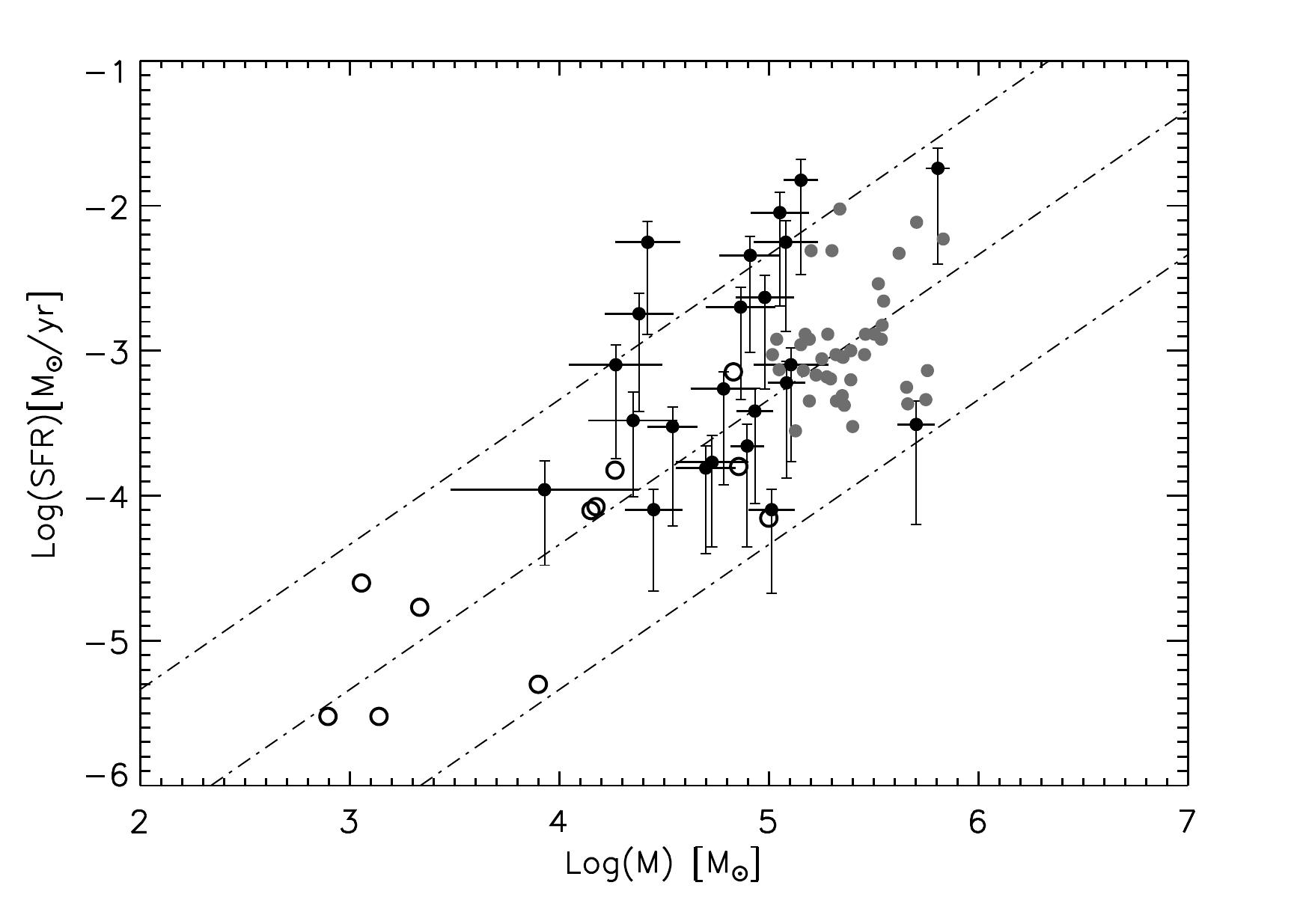} 
\caption{This figure shows a SFR-molecular-mass diagram for LMC and Milky Way molecular clouds.  The open circles and dashed lines are the same as those in Figure 1 of \citet{lada2012}.  The open circles correspond to cloud masses with extinction above A$_K$ = 0.1 mag (total cloud masses) and SFRs from Table 2 of \citet{lada2010}, which are based on YSO counting.  The dark circles correspond to the LMC SFRs and total cloud masses from \citet{magma} that are based on CO observations.  The gray dots are SFR and CO mass results from \citet{ngc300}.  The parallel dashed lines are linear relations that indicate constant fractions of dense gas, as described in \citet{lada2012}.}
\label{sfr_compare}	
\end{figure}

\subsubsection{Star Formation Rates}

Studies of local molecular clouds in the Milky Way by \citet{lada2009,lada2010} have established that there exists a large range in the star formation rates in molecular clouds independent of their sizes and total masses.  We see a similar phenomenon in the LMC 6\_6 tile molecular clouds, in that there is a wide range in the star formation rates of these clouds, which is shown in Figure~\ref{sfr}.  However, when we examine the overall star formation rates in the areas covered by both embedded cluster surveys, we find that the LMC 6\_6 tile is much more active than in the Solar neighborhood.  The SFR for the LMC 6\_6 tile can be calculated by summing all of the embedded cluster masses, then dividing by the age of the clusters (assumed to be 2 Myrs) and dividing by the area.  Using this method, we calculate the LMC 6\_6 tile SFR to be $\sim$5.93 $\times$ 10$^{-8}$M$_{\sun}$yr$^{-1}$pc$^{-2}$.  In comparison, the Milky Way SFR within 500 pc of the Sun is 3 $\times$ 10$^{-9}$M$_{\sun}$yr$^{-1}$pc$^{-2}$ \citep{ladalada}, which is $\sim$20 times less than the LMC 6\_6 tile SFR.  This is likely due to overall differences in the particular environments studied, particularly the enhanced amounts of star formation occurring in the 6\_6 tile, or could possibly be a metallicity effect.  The higher overall SFR in this region of the LMC may not be unexpected considering the higher embedded cluster and molecular cloud surface densities discussed in Section~\ref{surfacedensities}.

While the overall SFRs of the two regions vary considerably, we find in Figure~\ref{sfr_compare}, that they are consistent with each other when considered in the context of the SFR scaling law of \citet{lada2012}.  Figure~\ref{sfr_compare} shows a SFR-molecular-cloud-mass diagram for the LMC 6\_6 tile, local Milky Way molecular clouds, and the results from \citet{ngc300} for star-forming regions in NGC 300.  The open circles and dashed lines are the same as those in Figure 1 of \citet{lada2012}.  The open circles correspond to cloud masses with extinction above A$_K$ = 0.1 mag (total cloud masses) and SFRs from Table 2 of \citet{lada2010}, which are based on YSO counting.  The dark circles correspond to our LMC SFRs and total cloud masses from \citet{magma} that are based on CO observations.  The error bars show the effects of age assumptions on the SFR for the LMC 6\_6 tile molecular clouds, which is by far the largest potential source of errors.  The upper limit of the error bars shows the SFR if one assumes an age of 1 Myr and the lower limit shows the SFR if one assumes an age of 10 Myr. The gray dots are the results from \citet{ngc300} that are derived from CO observations and modeling using GALEX FUV, Spitzer 24$\mu$m, and H$\alpha$ observations.  The parallel dashed lines are linear relations that indicate the SFR scaling law with constant fractions of dense gas, as described in \citet{lada2012}.  As one can see, while the SFR of the LMC molecular clouds are generally greater than what is seen in the local Milky Way, the SFRs are generally consistent with the results from \citet{ngc300} and both their results and ours are consistent with the SFR scaling law presented in \citet{lada2012}.  This indicates that while the details of embedded cluster formation may vary between these environments, the overall process of cluster formation within molecular clouds may be universal.

\section{Conclusions}

We present initial results of the first systematic search for embedded star clusters beyond the Milky Way.  This is the first census of embedded clusters in a significantly different environment than the solar neighborhood.  We have explored a $\sim$1.65 deg$^2$ area in the LMC, which surrounds the well-known star-forming region 30 Doradus and contains the northern portion of the molecular ridge.  We have identified 67 embedded cluster candidates, 45 of which are newly discovered as clusters.  We have determined sizes, luminosities and masses for these embedded clusters, examined the star formation rates (SFRs) of the molecular clouds containing embedded clusters, and made a comparison between the LMC 6\_6 tile and the local Milky Way.  

Our results indicate that the characteristics of the embedded cluster population in GMCs in the LMC differ from what has been found within the environment of the solar neighborhood in several ways.  Specifically, we find that embedded clusters with luminosities similar to or greater than the Trapezium are common in the LMC.  In addition to being more luminous, the LMC embedded clusters are larger and more massive than those identified within 2.4 kpc of the Sun.  {We also find that the mass surface densities of the embedded clusters are approximately 40 times higher and molecular clouds are $\sim$3 times higher in the LMC than in the local environment, the SFR is $\sim$20 times higher, and the SFE is $\sim$10 times higher.  Although the overall SFR is much higher, the SFRs in LMC molecular clouds are consistent with the SFR scaling law from \citet{lada2012}.

While we find that the detailed properties of embedded clusters and star formation differ between the local Milky Way and the LMC 6\_6 tile environments, star formation ultimately appears to be a robust and universal process.

\section{Acknowledgements}
K.A.R. acknowledges support from a Space Grant Fellowship from the Florida Space Grant Consortium.  K.A.R. and E.L. acknowledge support from NSF Grant \#AST-1517724.  MRC acknowledges support from the UK's Science and Technology Facilities Council [grant number ST/M001008/1] and from the German Exchange Service.  We thank the Cambridge Astronomy Survey Unit (CASU) and the Wide Field Astronomy Unit (WFAU) in Edinburgh for providing calibrated data products under the support of the Science and Technology Facility Council (STFC) in the UK.  We also thank Dr. Charles Lada for useful discussions and comments on earlier versions of the manuscript.   Finally, we wish to thank the anonymous referee for constructive comments that improved this paper.

\end{document}